\documentclass[12pt]{iopart}
\usepackage{iopams}  
\usepackage{bm}
\usepackage{color}
\usepackage{graphicx}
\usepackage{cite}
\usepackage[normalem]{ulem}

\begin{document}

\title{Charge orders in organic charge-transfer salts}

\author{
Ryui Kaneko$^1$~\footnote{Present address: Institute for Solid State Physics, University of Tokyo,
5-1-5 Kashiwanoha, Kashiwa, Chiba 277-8581, Japan},
Luca F. Tocchio$^2$~\footnote{Present address: Institute for condensed matter physics 
and complex systems, DISAT, Politecnico di Torino, I-10129, Italy},
Roser Valent\'{i}$^1$, Federico Becca$^2$
}

\address{$^1$Institute for Theoretical Physics, Goethe University Frankfurt,
Max-von-Laue-Stra{\ss}e 1, D-60438 Frankfurt a.M., Germany}
\address{$^2$Democritos National Simulation Center, Istituto Officina dei
Materiali del CNR, and SISSA-International School for Advanced Studies, 
Via Bonomea 265, I-34136 Trieste, Italy}

\ead{rkaneko@issp.u-tokyo.ac.jp}

\begin{abstract}
Motivated by recent experimental suggestions of charge-order-driven ferroelectricity 
in organic charge-transfer salts, such as $\kappa$-(BEDT-TTF)$_2$Cu[N(CN)$_2$]Cl,
we investigate magnetic and charge-ordered phases that emerge in an extended 
two-orbital Hubbard model on the anisotropic triangular lattice at $3/4$ filling. 
This model takes into account the presence of two organic BEDT-TTF molecules, which 
form a dimer on each site of the lattice, and includes short-range intramolecular 
and intermolecular interactions and hoppings. By using variational wave functions 
and quantum Monte Carlo techniques, we find two polar states with charge 
disproportionation inside the dimer, hinting to ferroelectricity. These charge-ordered 
insulating phases are stabilized in the strongly correlated limit and their actual 
charge pattern is determined by the relative strength of intradimer to interdimer 
couplings. Our results suggest that ferroelectricity is not driven by magnetism, since 
these polar phases can be stabilized also without antiferromagnetic order and provide 
a possible microscopic explanation of the experimental observations. In addition, 
a conventional dimer-Mott state (with uniform density and antiferromagnetic order) and 
a nonpolar charge-ordered state (with charge-rich and charge-poor dimers forming a 
checkerboard pattern) can be stabilized in the strong-coupling regime. Finally, when 
electron-electron interactions are weak, metallic states appear, with either uniform 
charge distribution or a peculiar $12$-site periodicity that generates honeycomb-like 
charge order.
\end{abstract}

%\pacs{}
%\submitto{\NJP}  
%\maketitle
%\tableofcontents

\section{Introduction}\label{sec:intro}

Orbital, charge, and spin degrees of freedom are intertwined in correlated electron systems and 
the search for unconventional quantum phases emerging from the interplay of these degrees of 
freedom is a very active field of research in condensed-matter physics. In particular,  
multiferroicity~\cite{brink2008}, where magnetism and ferroelectricity coexist, has received a 
lot of attention in recent years. Conventionally, one can divide multiferroics into two groups: 
In type-I multiferroics, ferroelectricity and magnetism have different origins~\cite{khomskii2006}, 
while in type-II multiferroics, ferroelectricity occurs only in the magnetically ordered state, 
where, for example, it is induced by helical magnetic order in geometrically frustrated 
antiferromagnets~\cite{fiebig2005,katsura2005,mostovoy2006}. Recently, charge-order-driven 
ferroelectricity was proposed in organic charge transfer salts~\cite{lunkenheimer2015b},
such as $\kappa$-(ET)$_2$Cu[N(CN)$_2$]Cl~\cite{lunkenheimer2012} and 
$\alpha$-(ET)$_2$I$_3$~\cite{takahashi2006,yakushi2012,lunkenheimer2015a}, where ET stands 
for BEDT-TTF [bis(ethylenedithio)-tetrathiafulvalene]. In the former one, ferroelectric and 
antiferromagnetic order appear simultaneously and the emergence of charge order is still under 
debate~\cite{sedlmeier2012,tomic2013,lang2014}; instead, the latter one is nonmagnetic and 
ferroelectricity is observed in the presence of pronounced charge order. These observations 
have opened a debate about the nature and interplay of charge order, ferroelectricity and 
magnetism in these materials, which will be at the focus of this study.

The building blocks of the $\kappa$-(ET)$_2$X family (where X indicates a monovalent anion) are
strongly-coupled dimers of ET molecules forming a triangular lattice. These materials have been 
widely studied within the half-filled single-band Hubbard model on the anisotropic triangular 
lattice, where only a single orbital per dimer is retained~\cite{powell2011}. Indeed, because of 
the strong hybridization between ET molecules belonging to the same dimer, the gap between the 
bonding and anti-bonding orbitals is large; the former one is fully occupied, while the latter 
one is half filled, thus justifying a single-band picture. However, this coarse-grained approach 
cannot explain the possible emergence of ferroelectricity (or multiferroicity) in these materials, 
which has been suggested to arise from a charge disproportionation {\it within} each dimer. In this 
sense, the minimal model that could capture these features must include two molecular orbitals on 
each dimer and $3/4$ filling.

In the last decades there have been several attempts to obtain accurate values of the parameters
defining microscopic models that would capture the low-energy properties of charge-transfer salts.
The hopping integrals between the different molecular orbitals are found to significantly affect 
the nature of the ground states, as already reported in the first Hartree-Fock studies of correlated 
models for charge-transfer salts~\cite{kino1995a,kino1995b,kino1996}. These considerations motivated 
a revision of the first estimates of the hopping parameters, that were based on the extended 
H\"{u}ckel method~\cite{komatsu1996}, by means of density-functional calculations. Here, consistent
results for the hopping parameters of the $\kappa$-(ET)$_2$X family have been reported by three 
independent calculations~\cite{nakamura2009,kandpal2009,scriven2012}, while slightly different values
have been recently proposed~\cite{koretsune2014}. Besides the role of hopping parameters, the 
importance of Coulomb interactions between different molecules in organic systems has been intensively 
discussed within {\it ab-initio} calculations~\cite{nakamura2009,scriven2009,nakamura2012,shinaoka2012}. 
More recently, the analysis of various (low-energy) multiorbital models also points to the key role of 
intermolecular Coulomb interactions in order to describe complex phases relevant for charge-transfer 
salts. In particular, possible stripe and non-stripe charge orderings~\cite{seo2000,mori2016} and the 
mutual exclusion of ferroelectricity and magnetism~\cite{naka2010} have been discussed for various 
models with intermolecular interactions. In addition, the existence of a dipolar spin-liquid phase 
has been suggested~\cite{hotta2010,gomi2016} (possibly also explaining the dielectric anomaly in 
$\kappa$-(ET)$_2$Cu$_2$(CN)$_3$~\cite{jawad2010}). Furthermore, the two-orbital Hubbard model has 
been claimed to be relevant for the description of superconductivity in charge-transfer 
salts~\cite{gomes2016,silva2016,guterding2016a,guterding2016}, including its proximity to charge-ordered 
phases~\cite{sekine2013,watanabe2017}. In addition, spin and charge fluctuations near the metal-insulator 
transition in multiorbital models have been analyzed~\cite{sato2017}.

In this paper, we concentrate on the question of what kind of charge orderings are driven by 
competing Coulomb interactions and which is their relation to ferroelectricity and magnetism. 
By using variational Monte Carlo methods, we investigate the phase diagram of an extended two-orbital 
Hubbard model on the triangular lattice at $3/4$ filling. Our results show that there exist two polar 
charge-ordered insulating phases, where charge disproportionation occurs within the dimer, and one 
nonpolar charge-ordered phase, with charge disproportionation between different dimers. All these 
phases are present also in the absence of magnetic order, indicating that they are not driven by 
magnetism. When magnetism is also included in the variational wave functions, we find that it coexists 
with charge order. These results could explain the observed behavior in $\kappa$-(ET)$_2$Cu[N(CN)$_2$]Cl. 
On the contrary, magnetism is crucial to stabilize the uniform dimer-Mott insulator, which appears 
in a narrow region between the two polar phases. In this respect, it has been experimentally suggested 
that a transition between the dimer-Mott insulator and charge-ordered states is a common feature 
among organic systems~\cite{okazaki2013}. Finally, when intramolecular and intermolecular Coulomb 
interactions are small and similar in magnitude, a metallic phase emerges, featuring charge order in 
the form of an effective honeycomb-lattice superstructure.

The paper is organized as follows: In Sec.~\ref{sec:model_methods}, we present the extended 
two-band Hubbard model for the organic charge transfer salts and the variational Monte Carlo 
method to study the phase diagram at zero temperature. In Sec.~\ref{sec:results}, we show the
numerical results and discuss the nature of charge-ordered phases. Finally, in Sec.~\ref{sec:summary}, 
we draw our conclusions.

\section{Model and methods}\label{sec:model_methods}

\subsection{The extended two-orbital Hubbard model}

\begin{figure}
\centering
\includegraphics[width=.8\textwidth]{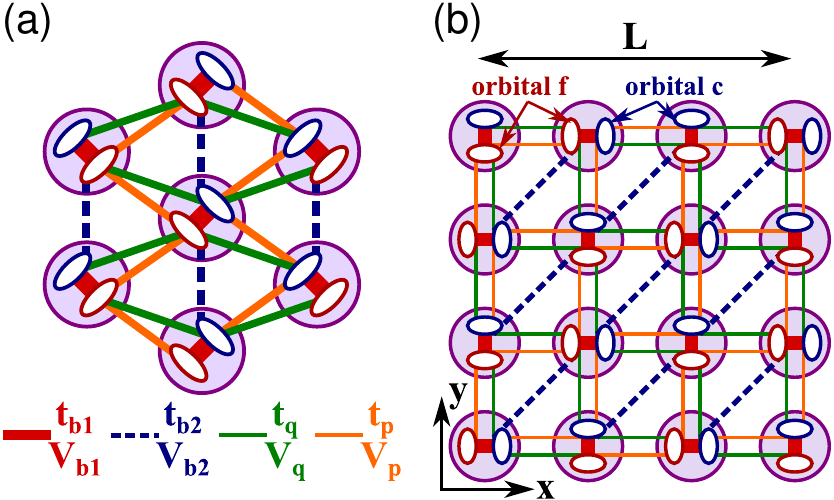}
\caption{(a) Dimer alignment in $\kappa$-(ET)$_2$X charge-transfer salts. (b) Equivalent square-lattice 
structure used in the calculations.}
\label{fig:lattice}
\end{figure}

In the following, we will consider a model in which every site (i.e., dimer) accommodates two orbitals
(hereinafter referred to as $c$ and $f$), one for each ET molecule. The original lattice is triangular, 
with hopping and interaction terms depicted in Fig.~\ref{fig:lattice}(a). An equivalent description is 
given by considering a two-orbital model on the square lattice, see Fig.~\ref{fig:lattice}(b). Here, 
we can define a partition in two sub-lattices ${\cal A}$ and ${\cal B}$, where the ET molecules form 
horizontal and vertical dimers, respectively. The full Hamiltonian, in this latter description, is given 
by:
\begin{equation}
{\cal H} = {\cal H}_t + {\cal H}_V + {\cal H}_U,
\label{eq:all}
\end{equation}
where
\begin{eqnarray}
\label{eq:kinetic} {\cal H}_t &=& t_{b1} \sum_{i,\sigma} c_{i,\sigma}^{\dagger} f_{i,\sigma}
                                + t_{b2} \sum_{i,\sigma} c_{i,\sigma}^{\dagger} f_{i+x+y,\sigma}
\nonumber \\
                               &+& t_{q} \sum_{i,\sigma} (c_{i,\sigma}^{\dagger} f_{i+x,\sigma}
                                      +c_{i,\sigma}^{\dagger} f_{i+y,\sigma})
\nonumber \\
                               &+& t_{p} \sum_{i\in {\cal A},\sigma} (c_{i,\sigma}^{\dagger} c_{i+x,\sigma}
                                                                     +c_{i,\sigma}^{\dagger} c_{i-y,\sigma}
                                                                     +f_{i,\sigma}^{\dagger} f_{i-x,\sigma}
                                                                     +f_{i,\sigma}^{\dagger} f_{i+y,\sigma})
                                 + \mathrm{h.c.},\\
{\cal H}_V &=& V_{b1} \sum_{i} n_{i}^{c} n_{i}^{f} + V_{b2} \sum_{i} n_{i}^{c} n_{i+x+y}^{f}
\nonumber \\
           &+& V_{q} \sum_{i} (n_{i}^{c} n_{i+x}^{f}+n_{i}^{c} n_{i+y}^{f})
\nonumber \\
           &+& V_{p} \sum_{i\in {\cal A}} (n_{i}^{c} n_{i+x}^{c}+n_{i}^{c} n_{i-y}^{c}
                                           +n_{i}^{f} n_{i-x}^{f}+n_{i}^{f} n_{i+y}^{f}), \\
{\cal H}_U &=& U \sum_{i} (n_{i,\uparrow}^{c} n_{i,\downarrow}^{c}+n_{i,\uparrow}^{f} n_{i,\downarrow}^{f}).
\end{eqnarray}
Here, $c^\dag_{i,\sigma}$ ($f^\dag_{i,\sigma}$) creates an electron with spin $\sigma$ on the $c$ ($f$) 
molecular orbital at site $i$; $n_{i}^{c} = \sum_{\sigma} n_{i,\sigma}^{c}$ 
($n_{i}^{f}=\sum_{\sigma}n_{i,\sigma}^{f}$) are the density operator for $c$ ($f$) electrons at site 
$i$. Hopping and interaction terms can be divided into those that connect $c$ and $f$ orbitals and
those that connect orbitals of the same kind, see Fig.~\ref{fig:lattice}(b). Belonging to the former class, 
there are terms connecting orbitals within the same dimer ($b_1$-type), along $x$ and $y$ nearest-neighbor 
sites ($q$-type), and along the $x=y$ diagonal ($b_2$-type); instead, $p$-type terms connect orbitals at 
nearest-neighbor sites and belong to the latter class. Accordingly, the noninteracting Hamiltonian 
${\cal H}_t$ contains four hopping parameters, i.e., $t_{b1}$, $t_{b2}$, $t_{q}$, and $t_{p}$. Similarly, 
the interacting Hamiltonian ${\cal H}_V$ contains four intermolecular Coulomb interactions, i.e., $V_{b1}$, 
$V_{b2}$, $V_{q}$, and $V_{p}$. Note that the translational symmetry and  the consequent partition between 
${\cal A}$ and ${\cal B}$ sub-lattices is only due to the presence of $p$-type terms. Finally, ${\cal H}_U$ 
describes the Hubbard-$U$ interaction on each molecule. Our calculations are performed on finite clusters 
of size $N_s=L^2$ (where on each site there are two molecules, i.e., orbitals), with periodic-antiperiodic boundary 
conditions on both directions. The filling factor is fixed to be $3/4$.

%%%%%%%%%%%%%%%%%%%%%%%%%%%%%%%%%%%%%%%%%

\begin{figure}
\centering
\includegraphics[width=.8\textwidth]{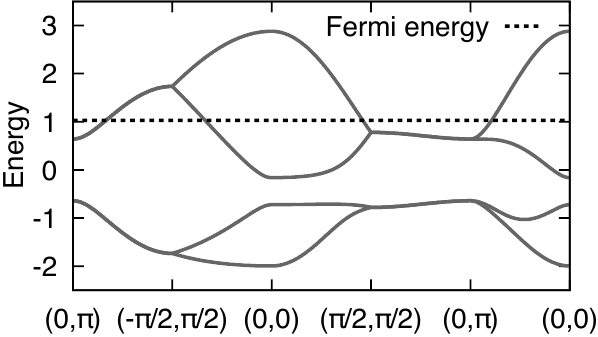}
\caption{Band structure for the set of parameters given in Eq.~(\ref{eq:hopping}). Four bands are present
because there are two orbitals per site and two inequivalent sites with vertical and horizontal dimers,
see Fig.~\ref{fig:lattice}.}
\label{fig:bands}
\end{figure}
%%%%%%%%%%%%%%%%%%%%%%%%%%%%%%%%%%%%%%%%%

As discussed in Ref.~\cite{guterding2016}, this two-orbital model reduces to the single-band Hubbard 
model (at half filling), when the intradimer hopping is very large (i.e., $t_{b1} \gg t_{b2},t_{p},t_{q}$). 
Furthermore, at $t_{b2}=0$ and $t_{p}=0$ (or $t_{q}=0$) the Hamiltonian reduces to the recently investigated
Hubbard model on the honeycomb lattice with anisotropic terms~\cite{kaneko2016b}.

In this work, we consider the following hopping parameters (in units of $t_{b1}$): 
\begin{equation}\label{eq:hopping}
 t_{b1} = 1,~
 t_{b2} = 0.359,~
 t_{p} = 0.539,~
 t_{q} = 0.221,
\end{equation}
which are based on the results obtained by density-functional-theory calculations on 
$\kappa$-(ET)$_2$Cu[N(CN)$_2$]Cl~\cite{guterding2016,private_micha}. The noninteracting band structure is 
reported in Fig.~\ref{fig:bands}. As far as the interaction terms are concerned, for realistic systems,
one expects $U$ to be the largest Coulomb repulsion term and $V_{b1}$ to be the second largest one, 
while $V_{b2}$ should be comparable to $V_{p}$ and $V_{q}$.

%%%%%%%%%%%%%%%%%%%%%%%%%%%%%%%%%%%%%%%%%
\begin{figure}
\centering
\includegraphics[width=.9\textwidth]{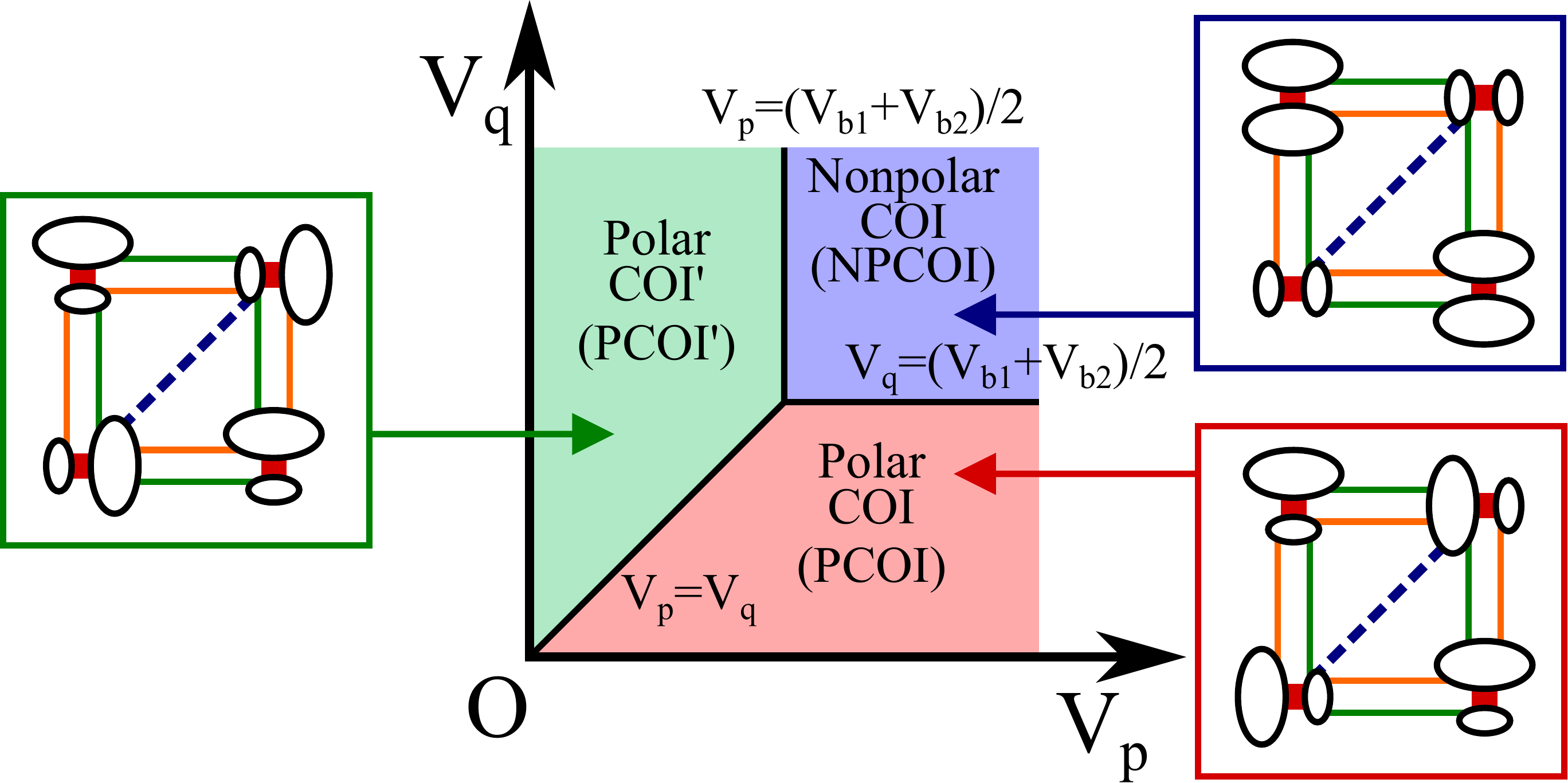}
\caption{Schematic phase diagram of model Eq.~(\ref{eq:all}) in the atomic limit ($t_{b1}=t_{b2}=t_{p}=t_{q}=0$), 
with two polar charge-ordered insulators (PCOI and PCOI$^\prime$) and one nonpolar charge-ordered insulator (NPCOI). 
Large ovals represent doubly occupied molecules, while small ovals represent singly-occupied molecules. The spin 
configurations on singly-occupied molecules have macroscopic degeneracy.}
\label{fig:Uinf_phase_diag}
\end{figure}
%%%%%%%%%%%%%%%%%%%%%%%%%%%%%%%%%%%%%%%%%

\subsection{The atomic limit}

We first discuss the possible ground states in the atomic limit, i.e., for $t_{b1}=t_{b2}=t_{p}=t_{q}=0$ at 
$3/4$ filling. If the only finite interaction is the intramolecular Hubbard-$U$ term, the ground state is highly 
degenerate, with all possible charge patterns having $N_s$ doubly-occupied molecules. A further degeneracy arises 
from the remaining $N_s$ molecules being singly-occupied, where any spin configuration gives the same energy. 
The charge degeneracy can be lifted by including the intermolecular interactions $V_{b1}$, $V_{b2}$, $V_{p}$, 
and $V_{q}$. We concentrate here on three particular relevant cases that show regular patterns of charge order 
(see Fig.~\ref{fig:Uinf_phase_diag}): Two of them are polar charge-order insulators (hereinafter denoted as PCOI 
and PCOI$^\prime$), since there is a charge disproportionation inside each dimer, and one is a nonpolar charge-order 
insulator (denoted as NPCOI), since the two molecules of the same dimer have the same amount of charge. Their 
energies per site (i.e., per dimer) can be easily evaluated in the atomic limit:
\begin{eqnarray}
\label{eq:polar1} E_{\rm polar} &=& E + V_q, \\
\label{eq:polar2} E_{\rm polar'} &=& E + V_p, \\
\label{eq:nonpol} E_{\rm nonpolar} &=& E + \frac{1}{2} (V_{b1}+V_{b2}),
\end{eqnarray}
where we defined:
\begin{equation}
E = U + 2V_{b1} + 4V_p + 4V_q + 2V_{b2}.
\end{equation}
The phase diagram in the $V_p{-}V_q$ plane is shown in Fig.~\ref{fig:Uinf_phase_diag}. Here, $V_{b1}$ 
and $V_{b2}$ only modify the phase boundaries between the polar and nonpolar charge-ordered phases.
The NPCOI appears when both $V_p$ and $V_q$ dominate over $V_{b1}$ and $V_{b2}$. Otherwise, the two polar 
states are stable and the competition between $V_p$ and $V_q$ determines the detailed charge pattern. 
All three phases are degenerate for $V_p=V_q=(V_{b1}+V_{b2})/2$.

\subsection{Variational wave functions}

Our numerical results are obtained by means of the variational Monte Carlo method, that is based on the 
definition of suitable wave functions that approximate the ground-state properties beyond perturbative 
approaches. We consider Jastrow-Slater wave functions~\cite{yokoyama1987,gros1988,capello2005,tocchio2016}, 
which are described as:
\begin{equation}\label{eq:wf}
|\Psi\rangle = {\cal J} |\Phi\rangle.
\end{equation}
Here, ${\cal J}$ is a long-range density-density Jastrow factor given by:
\begin{equation}
{\cal J} = \exp \left( -\frac{1}{2}\sum_{i,j,\alpha,\beta} v_{ij}^{\alpha\beta} n_{i}^{\alpha} n_{j}^{\beta} \right),
\end{equation}
where $n_i^{\alpha}$ is the electronic density at site $i$ and orbital $\alpha=c,f$, while $v_{ij}^{\alpha\beta}$ 
are translational invariant variational parameters, that are optimized by only imposing translational and inversion 
symmetry in the square lattice defined by $\bm{R}_i=(x_i,y_i)$. $|\Phi\rangle$ is a noninteracting fermionic state 
that is defined as the ground state of an auxiliary Hamiltonian with site-dependent chemical potentials and magnetic 
order parameters. Such a choice of an auxiliary Hamiltonian allows us to describe both charge and spin orders induced 
by the intermolecular Coulomb interactions~\cite{tocchio2014,kaneko2016a,kaneko2016b}. In particular, for the 
{\it insulating} states, we consider:
\begin{equation}
{\cal H}_{\rm ins} = {\cal H}_{t} + {\cal H}_{\rm COI} + {\cal H}_{\rm AF},
\end{equation}
where ${\cal H}_{t}$ is the kinetic part of Eq.~(\ref{eq:kinetic}) and
\begin{eqnarray}
\label{eq:insCO} {\cal H}_{\rm COI} &=& \sum_{i} e^{i\bm{Q}\cdot\bm{R}_i} (\mu^c n_{i}^{c} + \mu^f n_{i}^{f}), \\
\label{eq:insAF} {\cal H}_{\rm AF} &=& \sum_{i} [m_{i}^{c} (c^\dag_{i,\uparrow} c_{i,\downarrow} + 
c^\dag_{i,\downarrow} c_{i,\uparrow}) + m_{i}^{f} (f^\dag_{i,\uparrow} f_{i,\downarrow} +
f^\dag_{i,\downarrow} f_{i,\uparrow})].
\end{eqnarray}
Here, $\bm{Q}=(\pi,\pi)$ describes the NPCOI (with $\mu^c=\mu^f$) and the PCOI (with $\mu^c=-\mu^f$) phases 
of Fig.~\ref{fig:Uinf_phase_diag}, while $\bm{Q}=(0,0)$ (with $\mu^c=-\mu^f$) gives the PCOI$^\prime$ phase 
of Fig.~\ref{fig:Uinf_phase_diag}. We optimize the variational magnetic parameters at charge-rich and charge-poor 
molecular orbitals independently, according to the condition:
\begin{equation}\label{eq:chemical}
m_{i}^{\alpha} = \left \{
\begin{array}{ll}
m_1^{\alpha} & {\rm ~if~} e^{i\bm{Q}\cdot\bm{R}_i}\mu^{\alpha}<0, \\
m_2^{\alpha} & {\rm ~if~} e^{i\bm{Q}\cdot\bm{R}_i}\mu^{\alpha}>0,
\end{array}
\right .
\end{equation}
which implies that $m_1^{\alpha}$ and $m_2^{\alpha}$ are associated to the magnetization of the charge-rich and
charge-poor molecular orbitals on site $i$, respectively. In general, incommensurate magnetic order may 
coexist with commensurate charge order; however, this is beyond the scope of the present paper, and we restrict 
ourselves to commensurate (and collinear) magnetic order. Notice that, within our variational description based
upon an auxiliary Hamiltonian, it is particularly easy to consider nonmagnetic states, which can be 
described by taking $m_{i}^{c}=m_{i}^{f}=0$ in Eq.~(\ref{eq:insAF}). 

In order to describe {\it metallic} states, 
we consider the following auxiliary Hamiltonian:
\begin{equation}\label{eq:metCO}
{\cal H}_{\rm met} = {\cal H}_{t} + {\cal H}_{\rm COM},
\end{equation}
where
\begin{equation}
{\cal H}_{\rm COM} = \sum_{i} [\mu_{R(i)}^{c} n_{i}^{c} + \mu_{R(i)}^{f} n_{i}^{f}];
\end{equation}
here, the sublattice index $R(i)$ at position $\bm{R}_i$ is defined as:
\begin{equation} \label{eq:sublattice}
R(i) = {\rm mod}(x_i-y_i,6),
\end{equation}
which allows a $12$-sublattice charge ordering, since there are two orbitals on each site. We neglect in the 
calculation the possible presence of magnetic order in the metallic states. 

In order to exclude the presence of further ordered phases in the explored regions of the phase diagram, we 
have also employed unbiased wave functions, where different charge orderings can spontaneously emerge. 
In particular, we constructed a noninteracting wave function $|\Phi\rangle$ as the ground state of the 
tight-binding Hamiltonian with site-dependent chemical potentials $\mu_{i}^{c}$ and $\mu_{i}^{f}$:
\begin{equation}
{\cal H}_{\rm full} = {\cal H}_{t} + \sum_{i} (\mu_{i}^{c} n_{i}^{c} +
\mu_{i}^{f} n_{i}^{f}), \end{equation}
where the chemical potentials are variational parameters {\it independently} optimized for each site $i$ 
(several initial configurations of $\mu_{i}^{c}$ and $\mu_{i}^{f}$ have been chosen, in order to assess the 
possibility to remain stuck in local minima). Since the number of parameters to be optimized grows as $2N_s$, 
we considered this approach only for $N_s=36$. In this case, we have observed that the selected charge orderings
are consistent with the states described by the simpler approach above. In addition, we notice that charge 
order can be also generated by a translationally invariant Jastrow factor, without explicitly breaking the 
symmetry in the Slater determinant, as shown in Refs.~\cite{tocchio2014,kaneko2016a}. The advantage is that 
we do not need to assume {\it a priori} any type of charge ordering and, if long-range order exists, 
charge-ordered states should be selected by the optimization of the Jastrow factor. In general, for the 
chosen set of hoppings and interaction terms, we never found charge orders that cannot be captured by the 
previous parametrization of Eqs.~(\ref{eq:insCO}) and~(\ref{eq:metCO}).

Given the presence of the correlation term (i.e., the Jastrow factor), an analytical evaluation of the 
variational energy or of any correlation function is impossible on large sizes; nevertheless, a standard 
Monte Carlo sampling can be employed to obtain all the physical quantities with high accuracy.
 
%%%%%%%%%%%%%%%%%%%%%%%%%%%%%%%%%%%%%%%%%
\begin{figure}
\centering
\includegraphics[width=.9\textwidth]{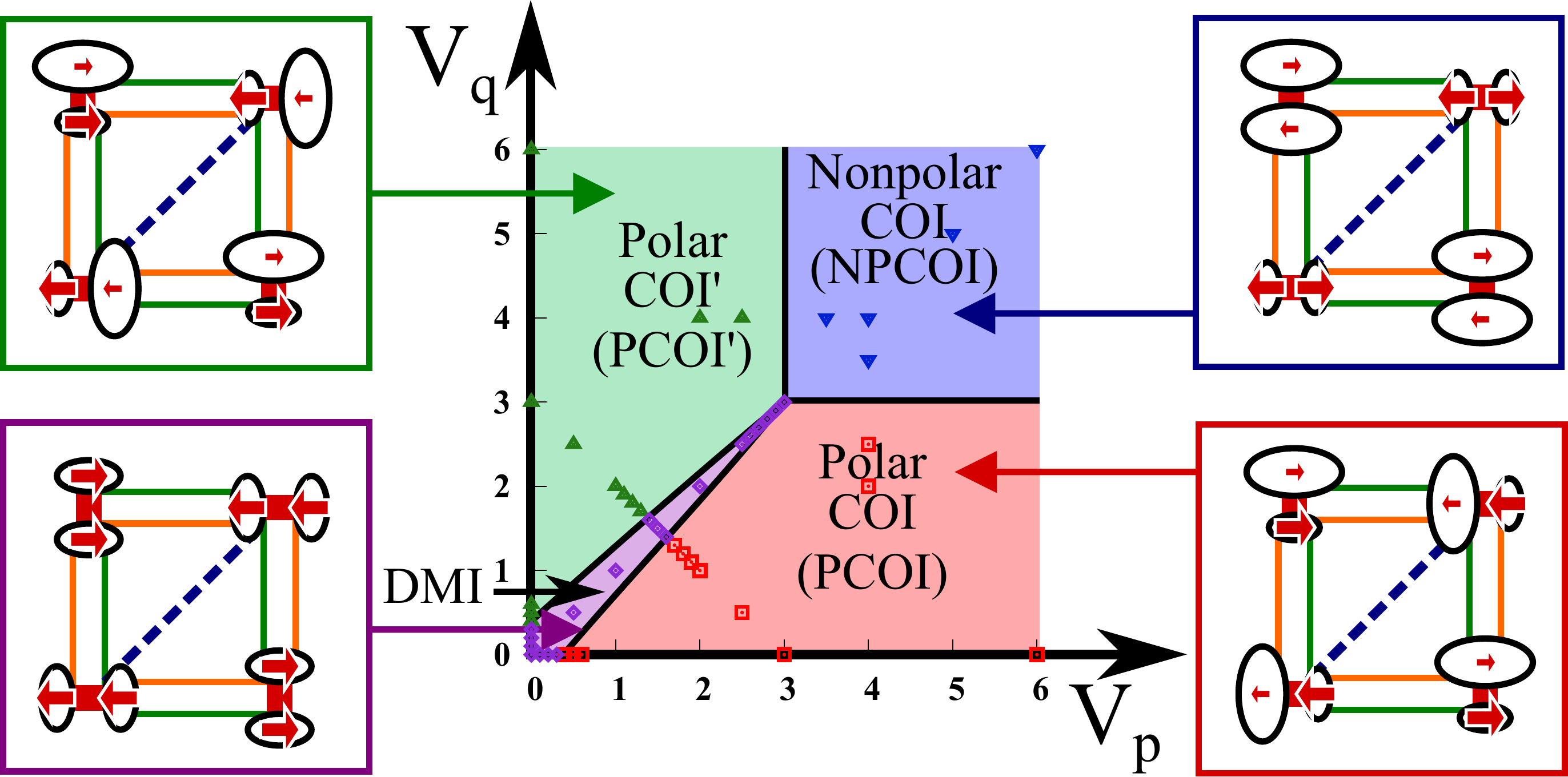}
\caption{Schematic phase diagram of model Eq.~(\ref{eq:all}) for large $U/t_{b1}$. In addition to the phases of 
the atomic limit (see Fig.~\ref{fig:Uinf_phase_diag}), the uniform dimer-Mott insulator (DMI) appears. Here, 
the points where we performed the variational calculations have been marked by green up-triangles (PCOI$^\prime$),
blue down-triangles (NPCOI), red squares (PCOI), and violet diamonds (DMI). Notice that finite nonzero hopping 
terms generate effective super-exchange couplings that stabilize antiferromagnetic order.}
\label{fig:UVlarge_phase_diag}
\end{figure}
%%%%%%%%%%%%%%%%%%%%%%%%%%%%%%%%%%%%%%%%%

\section{Results}\label{sec:results}

\subsection{Phase diagram for large $U$}

We now investigate how the hopping terms in Eq.~(\ref{eq:all}) modify the phase diagram obtained for the atomic 
limit, in the region $U \gg t_{b1}$. The schematic phase diagram is shown in Fig.~\ref{fig:UVlarge_phase_diag}. 
The three charge-ordered phases obtained in the atomic limit are stable also in the presence of finite values 
of the hopping terms; however, magnetic order is generated from virtual hopping processes involving charge-poor 
molecules, which form one-dimensional patterns in the lattice and are effectively half filled in all these phases. 
Antiferromagnetic correlations are then expected along these one-dimensional chains, which are formed by the bonds with
hopping $t_{b1}$ and $t_{b2}$ for the nonpolar state and by the bonds with hopping $t_p$ ($t_q$) for the PCOI$^\prime$
(PCOI). Therefore, in the nonpolar charge-ordered state, the two spins on the molecules of the same dimer have 
opposite orientations, thus implying that the dimer has no net magnetization. By contrast, the two polar states 
show ferromagnetic spin correlations within the dimers; here, each dimer contains one charge-rich and one 
charge-poor molecule, the magnetization being large in the latter one. Moreover, we observe long-range 
antiferromagnetic order of the magnetic moments of dimers. 

In addition to these three states, a uniform dimer-Mott insulator (DMI) intrudes between the polar phases. 
This correlated phase should appear when $U$ is much larger than all the $V$ terms, in the region where $V_p$ 
and $V_q$ are competing~\cite{kino1995a,kino1995b,kino1996}. Here, spin correlations are ferromagnetic within 
each dimer, since there is an average of three electrons per site: Two electrons have opposite spins and do not
contribute to the magnetic moment, which is fully due to the third electron that is delocalized between the 
two molecules. Similarly to the two polar charge-ordered states, also here the spins of the dimers possess 
long-range antiferromagnetic order. We find that the transitions between the DMI and the polar charge-ordered 
phases (PCOI and PCOI$^\prime$) are continuous (see below). Close to the boundaries between nonpolar and polar 
charge-ordered phases, the DMI state can also be stabilized; however its energy remains higher than the energies 
of the other phases, indicating that it is a metastable state.

%%%%%%%%%%%%%%%%%%%%%%%%%%%%%%%%%%%%%%%%%
\begin{figure}
\centering
\includegraphics[width=.8\textwidth]{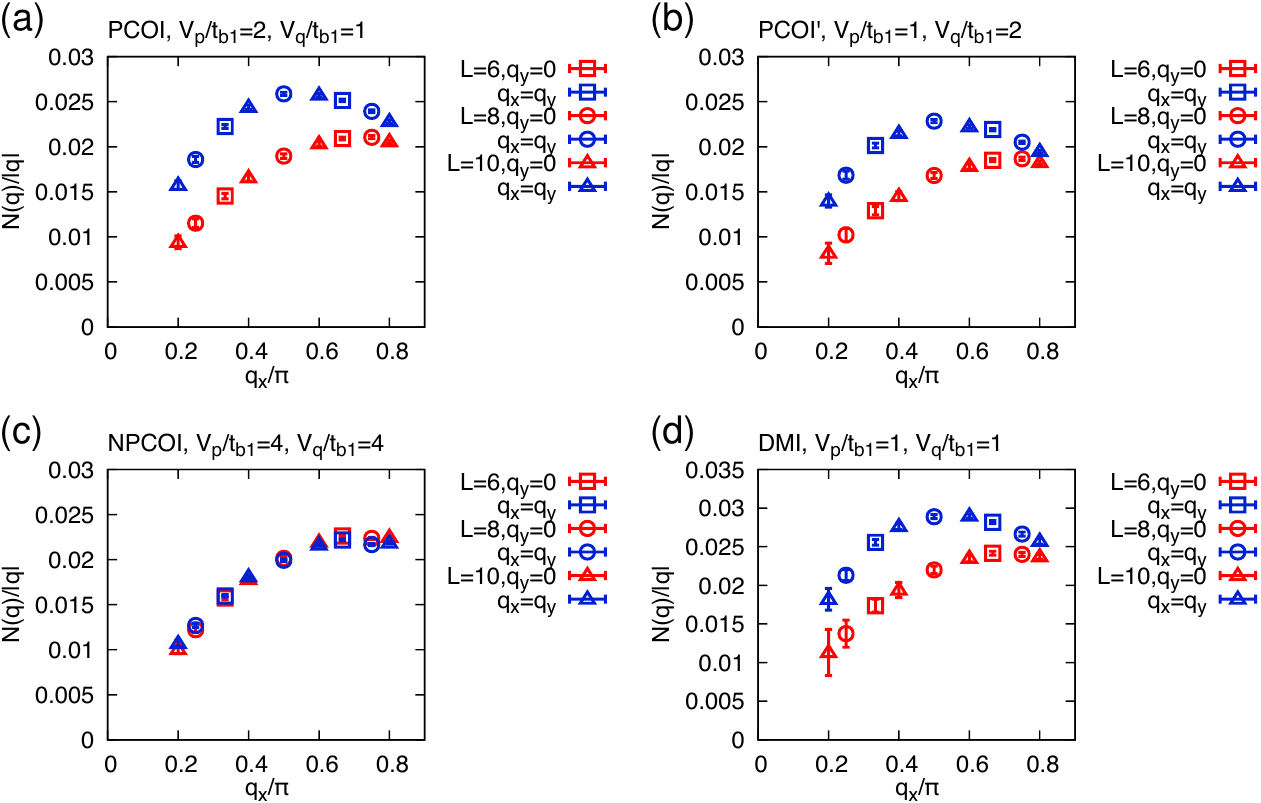}
\caption{Total charge structure factor for the four states in Fig.~\ref{fig:UVlarge_phase_diag}, divided by the 
momentum: $N(q)/|\bm{q}|$. $N(q) \propto |\bm{q}|^2$ for $|\bm{q}| \to 0$ in all cases, suggesting insulating 
behavior. Data are shown along the $q_y=0$ (red) and the $q_x=q_y$ (blue) lines in the Brillouin zone, for three 
lattice sizes: $L=6$ (squares), $L=8$ (circles), and $L=10$ (triangles).}
\label{fig:Nq_vs_q}
\end{figure}

\begin{figure}
\centering
\includegraphics[width=.8\textwidth]{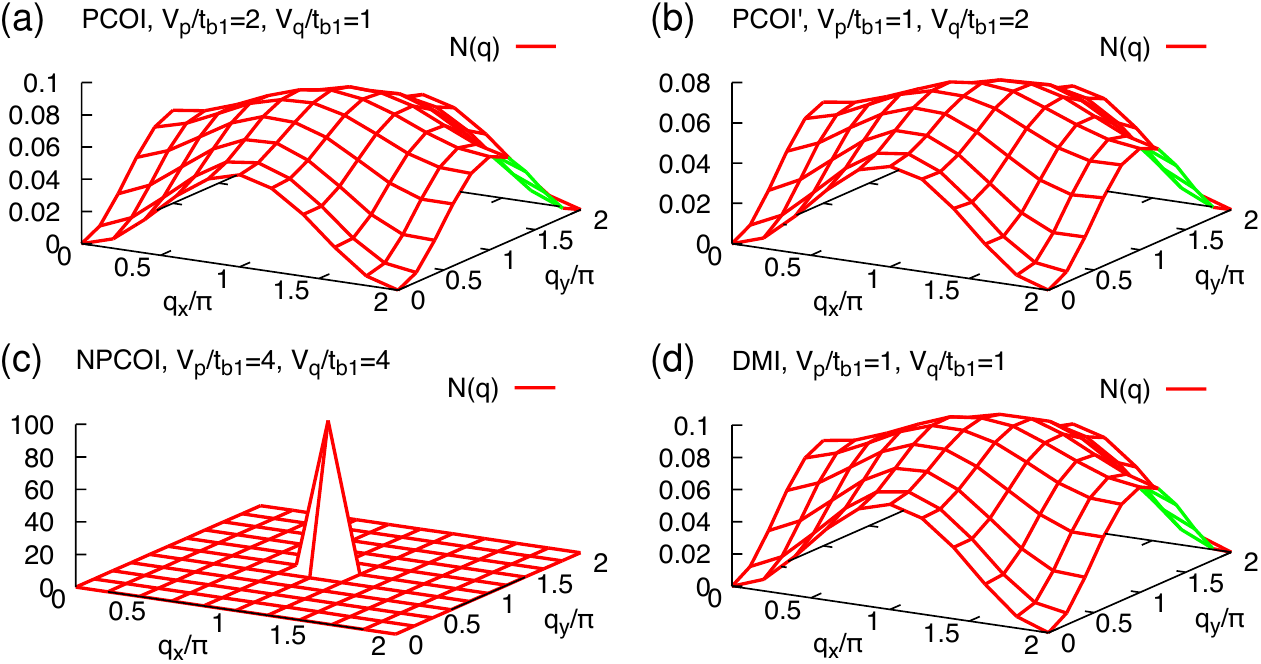}
\caption{Total charge structure factor $N(q)$, as a function of $\bm{q}$, for the four states in 
Fig.~\ref{fig:UVlarge_phase_diag}. Only the NPCOI state has a sharp peak at $\bm{Q}=(\pi,\pi)$, corresponding to 
interdimer charge disproportionation.}
\label{fig:Nq}
\end{figure}
%%%%%%%%%%%%%%%%%%%%%%%%%%%%%%%%%%%%%%%%%

In order to understand the nature of the charge properties of all the insulating phases, we calculate the total 
charge structure factor $N(q)$ and the charge-disproportionation structure factor $N_{\rm CD}(q)$, defined as:
\begin{eqnarray}
N(q) &=& \frac{1}{N_s} \sum_{i,j} \langle (n_{i}^{c} + n_{i}^{f}) (n_{j}^{c} + n_{j}^{f}) \rangle 
e^{i\bm{q}\cdot(\bm{R}_i-\bm{R}_j)}, \\
N_{\rm CD}(q) &=& \frac{1}{N_s} \sum_{i,j} \langle (n_{i}^{c} - n_{i}^{f}) (n_{j}^{c} - n_{j}^{f}) \rangle 
e^{i\bm{q}\cdot(\bm{R}_i-\bm{R}_j)},
\end{eqnarray}
where $\langle \dots \rangle$ indicates the expectation value over the variational wave function of Eq.~(\ref{eq:wf}).
Here, $N(q=0)$ is set to zero. The metallic or insulating character can be assessed by inspecting the small-$q$ limit 
of the total charge structure factor. Indeed, in the limit $|\bm{q}| \to 0$, $N(q) \propto |\bm{q}|$ for a metal, 
while $N(q) \propto |\bm{q}|^2$ for an insulator~\cite{feynman1954,tocchio2011}. In addition, charge order is indicated 
by the presence of a Bragg peak in $N(q)$ or $N_{\rm CD}(q)$. In the former case, charge order is characterized by 
charge-rich dimers on one sublattice and charge-poor dimers on the other one, while, in the latter case, charge 
disproportionation occurs within the dimers.

%%%%%%%%%%%%%%%%%%%%%%%%%%%%%%%%%%%%%%%%%
\begin{figure}
\centering
\includegraphics[width=.8\textwidth]{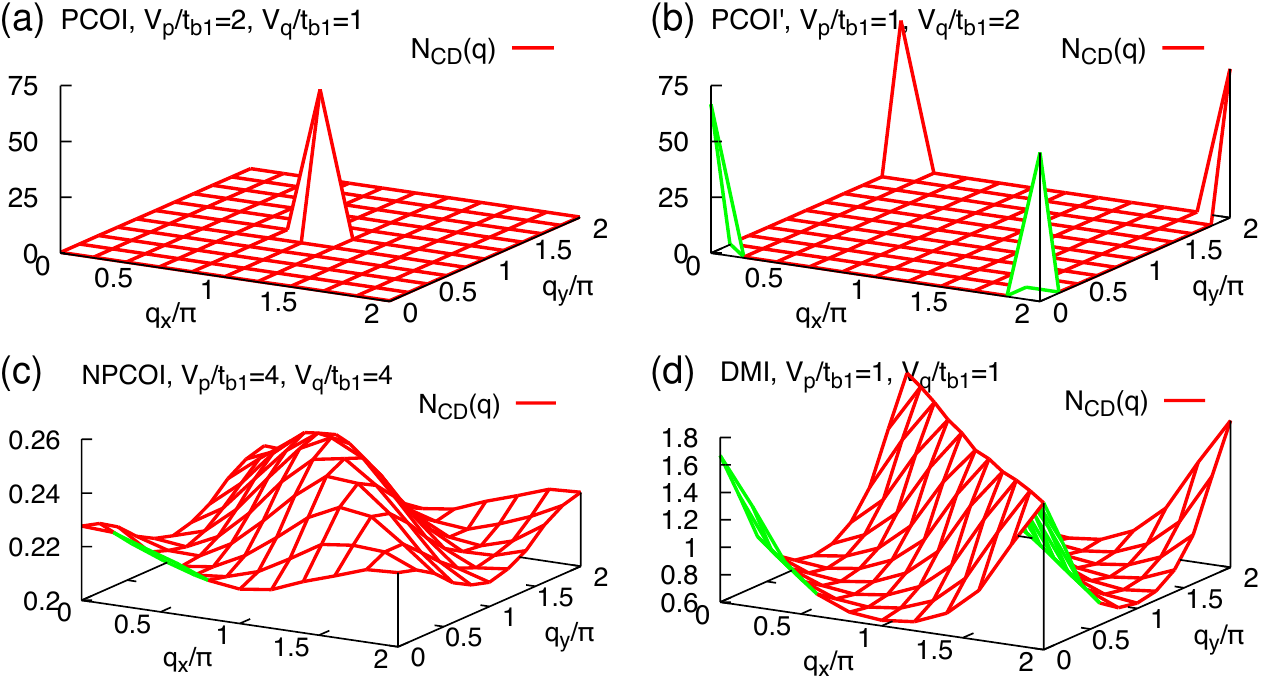}
\caption{Structure factor for charge disproportionation $N_{\rm CD}(q)$, as a function of $\bm{q}$, for the four 
states in Fig.~\ref{fig:UVlarge_phase_diag}. Only the PCOI and PCOI$^\prime$ states have a sharp peak at 
$\bm{Q}=(\pi,\pi)$ and at $\bm{Q}=(0,0)$, respectively, indicating charge disproportionation within the dimers.}
\label{fig:NCDq}
\end{figure}

\begin{figure}
\centering
\includegraphics[width=.8\textwidth]{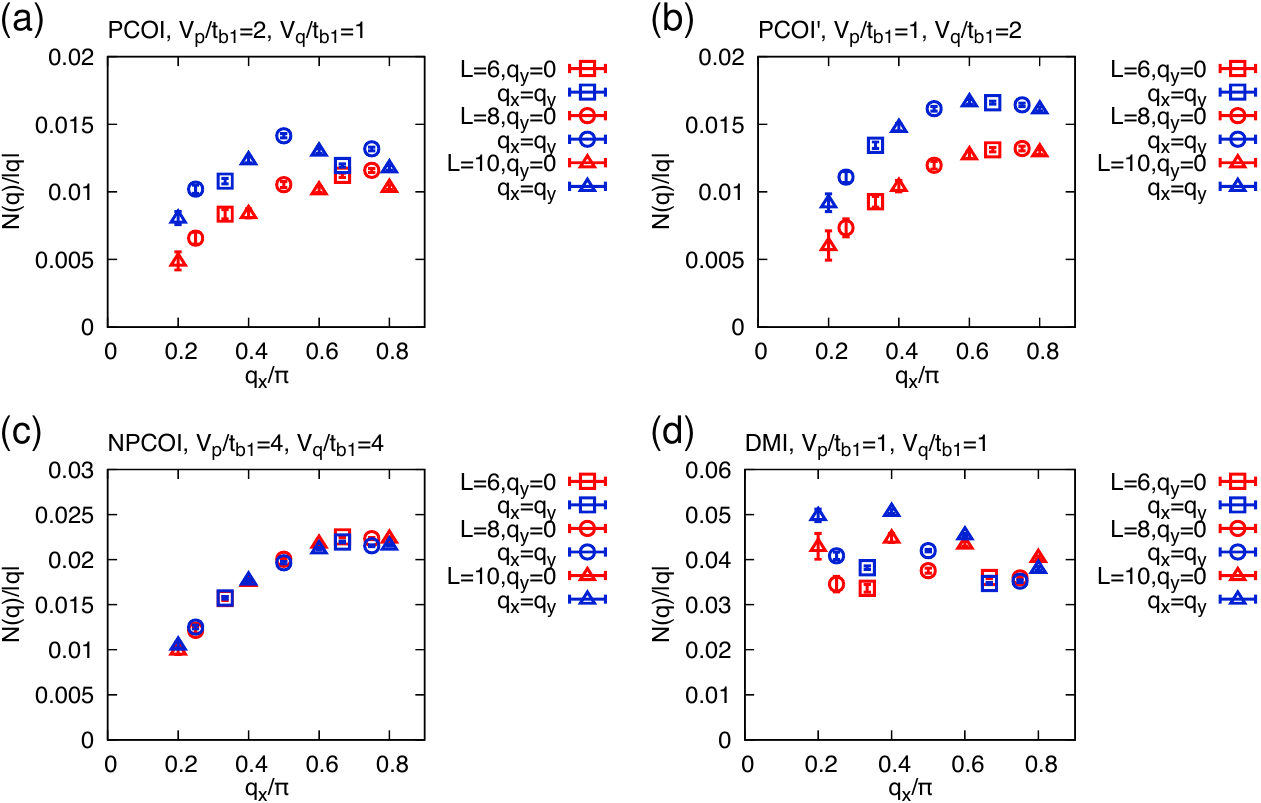}
\caption{The same as in Fig.~\ref{fig:Nq_vs_q}, but without including antiferromagnetic order in the variational 
wave functions.}
\label{fig:Nq_vs_q_NOAF}
\end{figure}
%%%%%%%%%%%%%%%%%%%%%%%%%%%%%%%%%%%%%%%%%

In the following, we fix the Coulomb interactions to $U/t_{b1}=10$, $V_{b1}/t_{b1}=4$, and $V_{b2}/t_{b1}=2$, and 
vary $V_p$ and $V_q$. Within this choice, in the atomic limit, the polar and nonpolar phases are degenerate for 
$V_p=V_q=3t_{b1}$. As shown in Fig.~\ref{fig:Nq_vs_q}, all the phases presented in the phase diagrams are insulating, 
since $N(q) \propto |\bm{q}|^2$ in the limit $|\bm{q}| \to 0$, both along the $q_y=0$ and the $q_x=q_y$ lines in 
reciprocal space. Then, each insulating phase can be fully characterized by $N(q)$ and $N_{\rm CD}(q)$, see 
Figs.~\ref{fig:Nq} and~\ref{fig:NCDq}. The DMI does not show any Bragg peak either in $N(q)$ or in $N_{\rm CD}(q)$ 
[Figs.~\ref{fig:Nq}(d) and ~\ref{fig:NCDq}(d)], suggesting that no long-range charge order occurs. The nonpolar 
charge-ordered state shows the Bragg peak at $\bm{Q}=(\pi,\pi)$ in $N(q)$ [Fig.~\ref{fig:Nq}(c)], but no sharp peaks 
in $N_{\rm CD}(q)$ [Fig.~\ref{fig:NCDq}(c)]. This implies that staggered charge disproportionation appears between 
different dimers, but not within the dimers. Finally, the polar charge-ordered states show the Bragg peak at 
$\bm{Q}=(0,0)$ (PCOI$^\prime$) or $\bm{Q}=(\pi,\pi)$ (PCOI) in $N_{\rm CD}(q)$ [Fig.~\ref{fig:NCDq} (a-b)],
but no sharp peaks in $N(q)$ [Fig.~\ref{fig:Nq} (a-b)]. This fact indicates charge disproportionation within the 
dimers, while the number of electrons in each dimer is constant. Each orbital has the same number of electrons 
at each site for $\bm{Q}=(0,0)$, while each orbital alternates between charge-rich and charge-poor configurations 
for $\bm{Q}=(\pi,\pi)$, see Fig.~\ref{fig:UVlarge_phase_diag}.

Remarkably, all polar and nonpolar phases can be stabilized within the variational approach also without considering 
magnetic order in the Slater determinant. By contrast, the DMI cannot be stabilized without including the
${\cal H}_{\rm AF}$ of Eq.~(\ref{eq:insAF}), see Fig.~\ref{fig:Nq_vs_q_NOAF}. The charge patterns are similar to 
the ones that have been obtained previously with the inclusion of magnetic order (not shown).

%%%%%%%%%%%%%%%%%%%%%%%%%%%%%%%%%%%%%%%%%
\begin{figure}
\centering
\includegraphics[width=.8\textwidth]{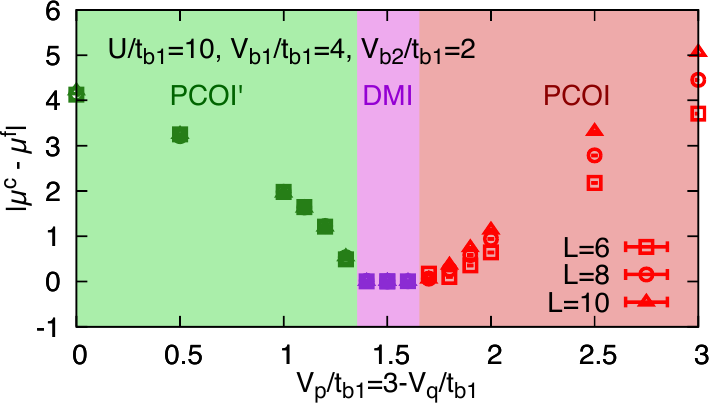}
\caption{Absolute value of the difference between the optimized orbital chemical potentials $|\mu^c-\mu^f|$ as a 
function of $V_p/t_{b1}$. The PCOI and PCOI$^\prime$ states are continuously connected to the DMI state. Data are 
shown for three lattice sizes $L=6$ (squares), $L=8$ (circles), and $L=10$ (triangles).}
\label{fig:mu_Vp_eq_3-Vq}
\end{figure}

\begin{figure}
\centering
\includegraphics[width=.8\textwidth]{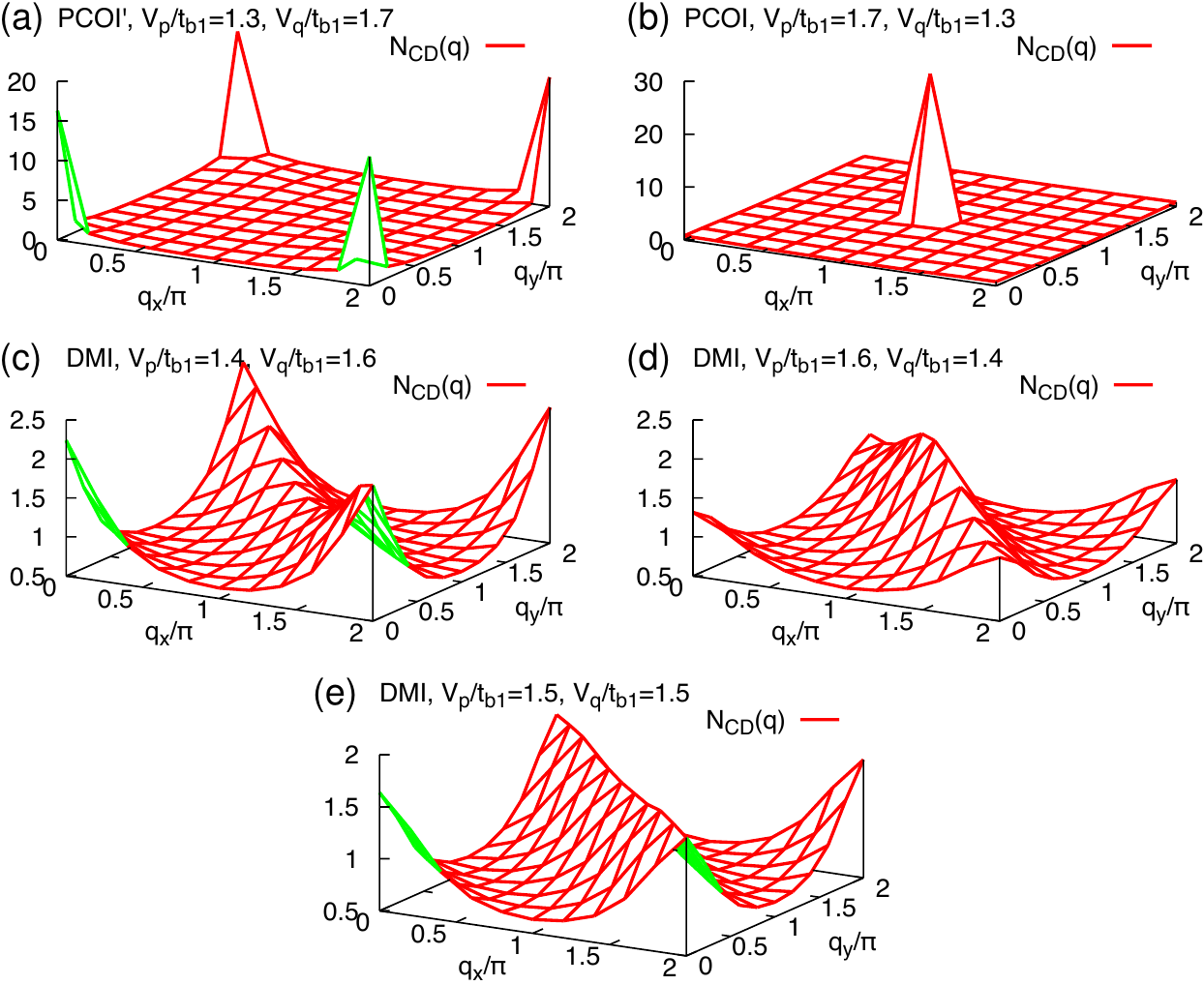}
\caption{Evolution of the charge-disproportionation structure factor $N_{\rm CD}(q)$ as a function of $\bm{q}$ 
along the $V_p+V_q=3t_{b1}$ line for the PCOI$^\prime$, PCOI, and DMI states in Fig.~\ref{fig:UVlarge_phase_diag}.}
\label{fig:NCDq_Vp_eq_3-Vq}
\end{figure}
%%%%%%%%%%%%%%%%%%%%%%%%%%%%%%%%%%%%%%%%%

\subsection{Competition between charge and magnetic orders}

We focus now on the interplay between charge and spin degrees of freedom near the boundary of the polar 
charge-ordered phases. In particular, we show the numerical results along the $V_p+V_q=3 t_{b1}$ line
(still fixing $U/t_{b1}=10$, $V_{b1}/t_{b1}=4$, and $V_{b2}/t_{b1}=2$), which crosses the two polar 
and the DMI phases. The absolute value of the difference between the optimized chemical potentials 
$|\mu^c-\mu^f|$ for orbitals $c$ and $f$ [see Eq.~(\ref{eq:insCO})] can be used as a diagnostic to 
detect polar and nonpolar states. As shown in Fig.~\ref{fig:mu_Vp_eq_3-Vq}, we find that $|\mu^c-\mu^f|$ 
is finite for $V_p/t_{b1} \lesssim 1.3$ and $V_p/t_{b1} \gtrsim 1.7$, while it vanishes in a narrow 
but finite region for $1.4\lesssim V_p/t_{b1}\lesssim 1.6$, indicating the existence of the DMI.
In addition, $|\mu^c-\mu^f|$ does not show any evidence of discontinuities at the transition points, 
strongly suggesting that the three phases are continuously connected. Indeed, near the phase boundary 
obtained in the atomic limit ($V_p/t_{b1}=1.5$) it is not possible to stabilize metastable wave functions 
with higher energies.

To further investigate the connection among these three phases, we calculate the charge-disproportionation
structure factor as function of $V_p$, as shown in Fig.~\ref{fig:NCDq_Vp_eq_3-Vq}. When $V_q$ ($V_p$) is 
sufficiently large, $N_{\rm CD}(q)$ shows a sharp peak at $\bm{Q}=(0,0)$ ($\bm{Q}=(\pi,\pi)$)
[Fig.~\ref{fig:NCDq_Vp_eq_3-Vq}(a) and (b), respectively]. By contrast, when $V_p \approx V_q$, 
$N_{\rm CD}(q)$ shows a broad crest along the $q_x+q_y=2\pi$ direction [Fig.~\ref{fig:NCDq_Vp_eq_3-Vq}(c-e)]. 
Importantly, there are no divergences in the thermodynamic limit, since the crest remains finite when 
increasing the size of the cluster. The peculiar one-dimensional-like shape of $N_{\rm CD}(q)$ might be 
understood in the following simple way: For $V_p \approx V_q$, the emergence of charge order is controlled 
only by $V_{b1}$ and $V_{b2}$, which define diagonal chains in the lattice, see Fig.~\ref{fig:lattice}. 
It is then natural to expect that correlations do not show any dependence on the transverse direction.

%%%%%%%%%%%%%%%%%%%%%%%%%%%%%%%%%%%%%%%%%
\begin{figure}
\centering
\includegraphics[width=.8\textwidth]{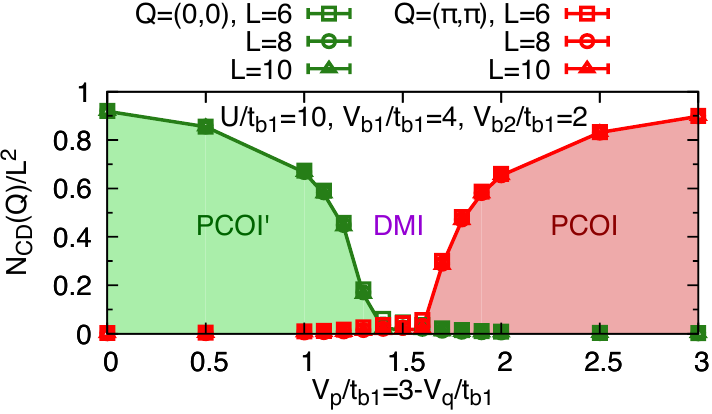}
\caption{Evolution of $N_{\rm CD}(Q)$ for $\bm{Q}=(0,0)$ (green) and $\bm{Q}=(\pi,\pi)$ (red), divided by 
$L^2$, as a function of $V_p/t_{b1}=3-V_q/t_{b1}$. The DMI phase is stabilized in a narrow region where both peaks 
in $N_{\rm CD}(Q)$ do not diverge with the system size. Data are shown for three lattice sizes: $L=6$ (squares), 
$L=8$ (circles), and $L=10$ (triangles).}
\label{fig:NCDqdivNs_Vp_eq_3-Vq}
\end{figure}

\begin{figure}
\centering
\includegraphics[width=.8\textwidth]{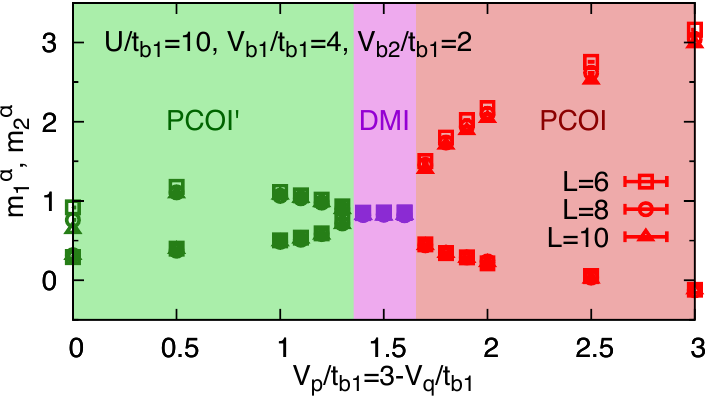}
\caption{Optimized magnetic order parameters for the charge-rich ($m_1^{\alpha}$) and for the charge-poor 
($m_2^{\alpha}$) molecules, as a function of $V_p/t_{b1}$. $m_1^{\alpha}<m_2^{\alpha}$ in the charge-ordered 
phases, while they become equal in the uniform DMI. Data are shown for three lattice sizes: 
$L=6$ (squares), $L=8$ (circles), and $L=10$ (triangles).}
\label{fig:mag_Vp_eq_3-Vq}
\end{figure}
%%%%%%%%%%%%%%%%%%%%%%%%%%%%%%%%%%%%%%%%%

The absence of charge disproportionation for $1.4\lesssim V_p/t_{b1} \lesssim 1.6$ is clearly demonstrated
by performing the size scaling of $N_{\rm CD}(Q)/L^2$ for $L \to \infty$. The results are reported in 
Fig.~\ref{fig:NCDqdivNs_Vp_eq_3-Vq}. For $V_p/t_{b1} \lesssim 1.3$ and $V_p/t_{b1} \gtrsim 1.7$, we have 
that $N_{\rm CD}(Q)/L^2$ is finite for $\bm{Q}=(0,0)$ and $\bm{Q}=(\pi,\pi)$, respectively. Instead, for 
$1.4\lesssim V_p/t\lesssim 1.6$ this quantity goes to zero for $L \to \infty$, suggesting that the DMI
is stable in this region.

Even if all the charge-ordered phases are present also in the absence of magnetic order, all of them are 
found to possess stable magnetic order when this possibility is included in the variational state, as shown 
in Fig.~\ref{fig:mag_Vp_eq_3-Vq}. The DMI shows the same absolute value of the magnetic moment for the 
orbitals $c$ and $f$, as expected for a charge uniform state. When the intersite Coulomb interactions 
become anisotropic (i.e., $|V_p-V_q|>0$), magnetic orders for the orbitals $c$ and $f$ start to deviate. 
This is due to the fact that the charge-rich (charge-poor) molecular orbitals possess a smaller (larger) magnetic moment 
in the polar charge-ordered phases. Notice that the PCOI state has a larger magnetic order than the PCOI$^\prime$ 
one. This may be due to the anisotropy in the hopping terms. Indeed, in the PCOI state the singly-occupied 
molecules are connected by $t_q$, while in the PCOI$^\prime$ one they are connected by $t_p$; since $t_p>t_q$ 
[see Eq.~(\ref{eq:hopping})] the latter case has more charge fluctuations (i.e., it is closer to a metal-insulator 
transition), thus implying smaller magnetic moments.

%%%%%%%%%%%%%%%%%%%%%%%%%%%%%%%%%%%%%%%%%
\begin{figure}
\centering
\includegraphics[width=.8\textwidth]{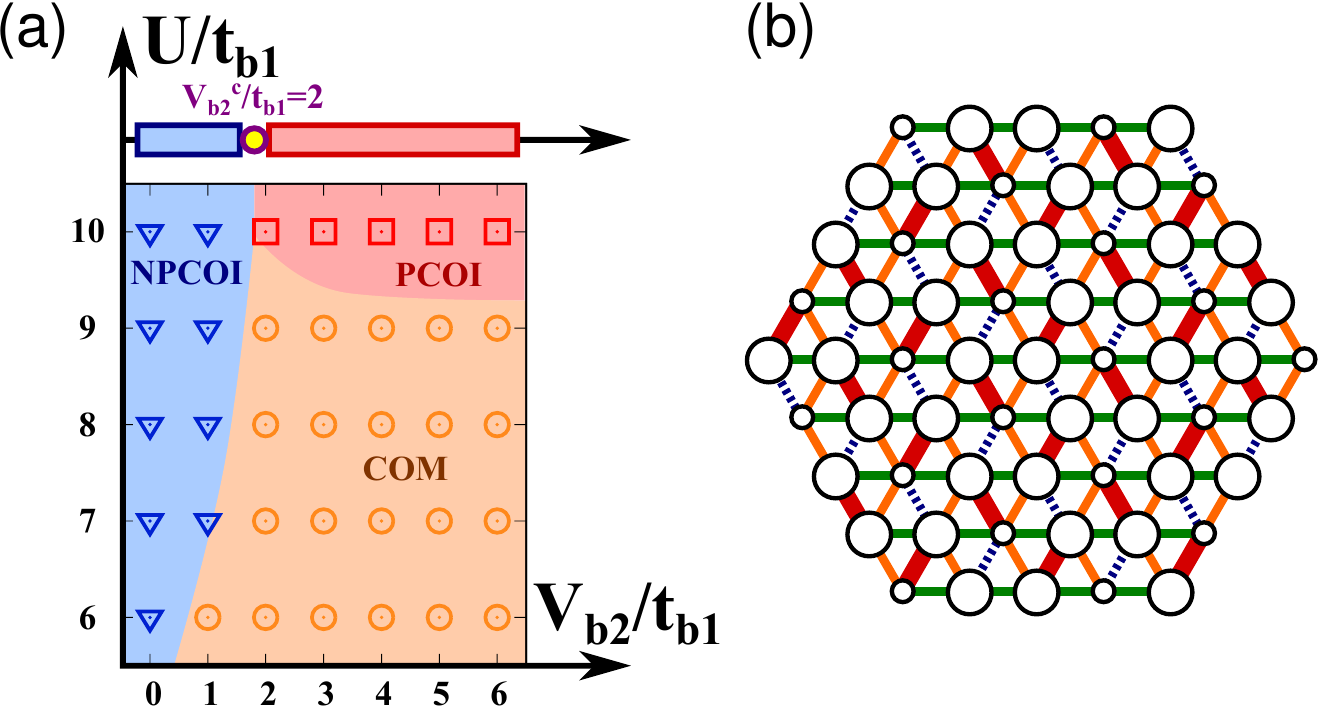}
\caption{(a) Phase diagram in the $U{-}V_{b2}$ plane, obtained by fixing $V_{b1}/t_{b1}=4$, $V_p/t_{b1}=3.5$, 
and $V_q/t_{b1}=3$. Three phases are present: the nonpolar charge-ordered insulator (NPCOI), the polar charge 
ordered insulator (PCOI), and a $12$-sublattice charge ordered metal (COM). The location of the transition 
between the NPCOI and PCOI phases for large $U$ is in agreement with the atomic limit, see Fig.~\ref{fig:Uinf_phase_diag}.
(b) Schematic charge configuration of the $12$-sublattice charge-ordered metal.}
\label{fig:COMetal_phase_diag}
\end{figure}

\begin{figure}
\centering
\includegraphics[width=.9\textwidth]{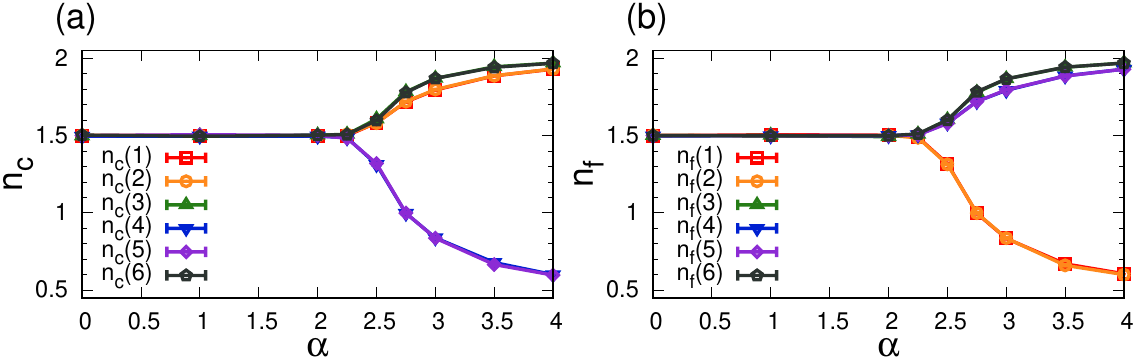}
\caption{Electron density in each of the six sublattices defined in Eq.~(\ref{eq:sublattice}), for orbitals 
$c$ and $f$, as a function of $\alpha=V_{b1}/t_{b1}$, which controls the strength of the intermolecular Coulomb 
interactions. Data are shown on the $L=12$ lattice size.}
\label{fig:COMetal_stability}
\end{figure}
%%%%%%%%%%%%%%%%%%%%%%%%%%%%%%%%%%%%%%%%%

\subsection{$12$-site ordered metallic phase}

Finally, we focus our attention on the charge-ordered metal (COM) that appears for small values of $U/t_{b1}$.
Therefore, we now fix $V_{b1}/t_{b1}=4$, $V_p/t_{b1}=3.5$, and $V_q/t_{b1}=3$ and vary $U/t_{b1}$ and $V_{b2}/t_{b1}$.
The phase diagram is shown in Fig.~\ref{fig:COMetal_phase_diag}(a). Here, a large metallic phase, with honeycomb-like 
charge ordering, appears, for relatively small values of the intramolecular interaction. This charge-ordered pattern 
is similar to the three-sublattice one~\cite{tocchio2014,hotta2006,fevrier2015}, which has been stabilized on the 
triangular lattice for intermediate values of the nearest-neighbor interaction. In our case, the periodicity is 
extended to $12$ sites due to the anisotropy of the parameters.

The emergence of the honeycomb-like COM can be easily understood when all the bonds (of $b_1$-, $b_2$-, $p$-, and 
$q$-type) are equivalent. In this ``isotropic'' limit, when considering {\it each} molecule as an independent site, 
the underlying lattice becomes triangular, see Fig.~\ref{fig:COMetal_phase_diag}(b). In this case, by decreasing 
the intramolecular Coulomb interaction $U$, there is an insulator to metal transition, with a metallic phase below 
the critical point. Moreover, since intermolecular Coulomb interactions screen the actual value of $U$, the metallic 
phase is even more stable when the $V$'s are present in the model. However, the presence of intermolecular Coulomb 
interactions leads to a spontaneous symmetry breaking in the translational symmetry and to charge disproportionation, 
that on the triangular lattice it is natural to assume with a three-sublattice ordering $A{-}B{-}C$. For an average 
occupation per site (i.e., molecule) $n=3/2$, the only possible choice to minimize the energy loss due to the 
intermolecular interactions is then to reduce the electron occupation on one sublattice and increase it on the other 
two (the limiting case being $n_A=0.5$ and $n_B=n_C=2$).

%%%%%%%%%%%%%%%%%%%%%%%%%%%%%%%%%%%%%%%%%
\begin{figure}
\centering
\includegraphics[width=.8\textwidth]{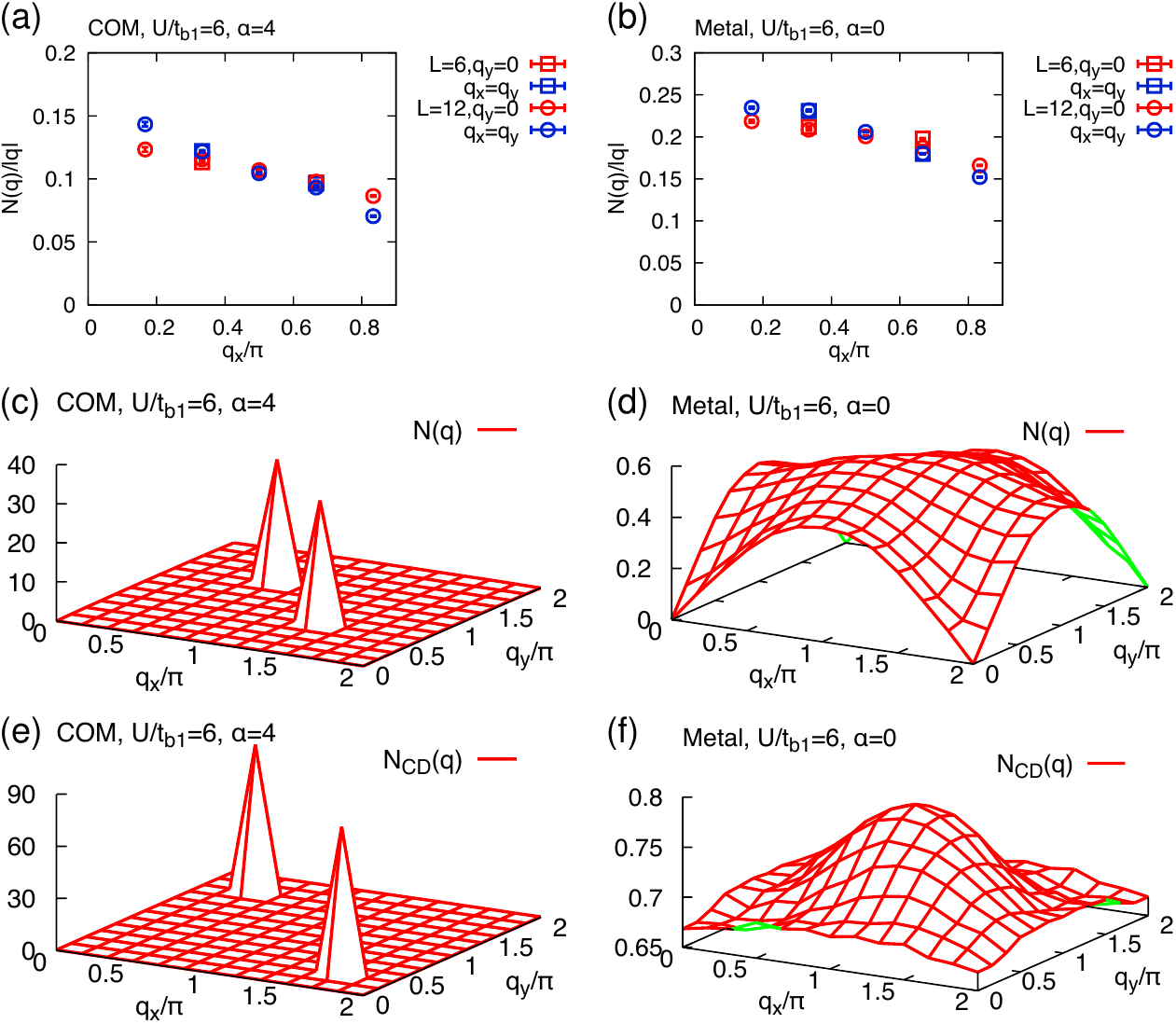}
\caption{Upper panels: Charge structure factor $N(q)$, divided by $|\bm{q}|$, for the COM phase (a) and for the 
uniform metallic phase (b). Data are shown along the $q_y=0$ (red) and the $q_x=q_y$ (blue) lines in reciprocal 
space, for $L=6$ (squares) and $L=12$ (circles). Middle panels: Charge structure factor $N(q)$ as a function of 
$\bm{q}$, for the COM phase (c) and for the uniform metallic phase (d). Lower panels: Structure factor for charge 
disproportionation $N_{\rm CD}(q)$ as a function of $\bm{q}$, for the COM phase (e) and for the uniform metallic 
phase (f).}
\label{fig:COMvsMetal}
\end{figure}
%%%%%%%%%%%%%%%%%%%%%%%%%%%%%%%%%%%%%%%%%

We investigate now the stability of the COM against a normal metal when decreasing the intermolecular interactions. 
In this case, we vary all the Coulomb terms together, taking $V_{b1}/t_{b1}=\alpha$, $V_{b2}/t_{b1}=0.5 \alpha$, 
$V_p/t_{b1}=0.875 \alpha$, and $V_q/t_{b1}=0.75 \alpha$, while the intramolecular interaction is fixed to $U/t_{b1}=6$.
As shown in Fig.~\ref{fig:COMetal_stability}, for $\alpha \lesssim 2$ the ground state is found to be a uniform 
metal, with no charge disproportionation. Charge order appears for $\alpha \approx 2.5$ and is characterized by the 
rich-rich-poor pattern. A direct comparison between the COM at $\alpha=4$ and the uniform metallic phase is presented 
in Fig.~\ref{fig:COMvsMetal}. Both phases are indeed metallic, since $N(q) \propto |\bm{q}|$ for small momenta 
[Figs.~\ref{fig:COMvsMetal}(a),(b)]. On the contrary, the formation of charge order in the COM phase, is signaled 
by the appearance of strong peaks in both the total charge structure factor $N(q)$ and the structure factor for 
charge disproportionation $N_{\rm CD}(q)$, corresponding to the real-space configuration illustrated in 
Fig.~\ref{fig:COMetal_phase_diag}(b).

\section{Summary and conclusions}\label{sec:summary}

By using variational wave functions and quantum Monte Carlo techniques, we have investigated the ground-state phase 
diagram of an extended two-orbital Hubbard model at $3/4$ filling on the anisotropic triangular lattice, which is
relevant for the $\kappa$-(ET)$_2$X family of organic charge-transfer salts. As a representative example, we have 
chosen the hopping parameters that correspond to $\kappa$-(ET)$_2$Cu[N(CN)$_2$]Cl and varied the interaction terms.
For large values of the intramolecular repulsion $U$ and by varying the strength of the competing intermolecular 
Coulomb interactions, we stabilize two polar and one nonpolar charge-ordered insulating phases, as well as a uniform 
dimer-Mott insulator. All these phases posses magnetic order, which is mainly determined by the behavior of the 
spins at the charge-poor molecules (that are effectively at half filling). We have also found that the dimer-Mott 
insulator is continuously connected to the two polar charge-ordered states: When the anisotropy between the 
intersite Coulomb interactions $V_p$ and $V_q$ goes to zero, the Bragg peaks of the two polar phases melt and form 
a one-dimensional-like structure. For smaller values of the intramolecular interaction $U$, we find a charge-ordered 
metal, that is similar to the three-sublattice (rich-rich-poor) charge order on the triangular lattice; however, 
the anisotropy in the intermolecular parameters modify the period of the charge ordering to a $12$-sublattice 
structure. Although charge-ordered metals in ET organic compounds often show a stripe-like charge 
pattern~\cite{watanabe2003,kakiuchi2007}, the observation of the COM phase would be an intriguing proof for the 
possibility of stabilizing nontrivial charge orders in metallic phases.

In organic charge transfer salts, the size of the intermolecular Coulomb interactions is expected to be larger when 
the molecules are closer. In this respect, it is plausible to assume that $V_{b1}$ is the stronger intermolecular 
Coulomb interaction and that $V_p \gtrsim V_q$, see Fig.~\ref{fig:lattice}. In addition to the fact that the strongest 
Coulomb interaction is the intramolecular one $U$, most of the compounds should be located at the border between 
the PCOI and the DMI phases. Since the two phases are continuously connected, a small amount of anisotropy 
$V_p \gtrsim V_q$ will lead to a weak charge order, as shown for example in Fig.~\ref{fig:mu_Vp_eq_3-Vq}; this fact 
may explain the difficulty in finding stable charge ordering in $\kappa$-(ET)$_2$Cu[N(CN)$_2$]Cl. Nevertheless, our 
results indicate that the PCOI phase is polarized, suggesting that charge order is the correct mechanism to induce a 
finite polarization. Moreover, we observe that magnetism coexists with electronic polarization, as observed in 
experiments, even if it is not the driving mechanism for it, since polarization occurs also in the absence of 
magnetic order. In this respect, our study shows that ferroelectricity in organic charge-transfer salts is not driven 
by magnetism.

Finally, we would like to conclude by mentioning that superconducting pairing correlations (with unconventional pairing 
symmetries) may be enhanced close to charge-ordered phases in multiorbital Hubbard models~\cite{watanabe2017}.
Investigating possible superconductivity (also coexisting with charge ordering) is left for future studies.

\ack

The authors would like to thank M. Altmeyer, C. Gros, D. Guterding, A.J. Kim, and S.M. Winter for fruitful discussions. R.K. and R.V. acknowledge 
the support of the German Science Foundation (Deutsche Forschungsgemeinschaft) through Grant No.\ SFB/TR49.

%%%%%%%%%%%%%%%%%%%%%%%%%%%%%%%%%%%%%%%%%
\section*{References}
%%%%%%%%%%%%%%%%%%%%%%%%%%%%%%%%%%%%%%%%%


\begin{thebibliography}{99}
\bibitem{brink2008} J. van den Brink and D.I. Khomskii, J. Phys.: Condens. Matter {\bf 20}, 434217 (2008).
\bibitem{khomskii2006} D.I. Khomskii, J. Magn. Magn. Mater. {\bf 306}, 1 (2006).
\bibitem{fiebig2005} M. Fiebig, J. Phys. D: Appl. Phys. {\bf 38}, R123 (2005).
\bibitem{katsura2005} H. Katsura, N. Nagaosa, and A.V. Balatsky, Phys. Rev. Lett. {\bf 95}, 057205 (2005).
\bibitem{mostovoy2006} M. Mostovoy, Phys. Rev. Lett. {\bf 96}, 067601 (2006).
\bibitem{lunkenheimer2015b} P. Lunkenheimer and A. Loidl, J. Phys.: Condens. Matter {\bf 27}, 373001 (2015).
\bibitem{lunkenheimer2012} P. Lunkenheimer, J. M\"{u}ller, S. Krohns, F. Schrettle, A. Loidl, B. Hartmann,
   R. Rommel, M. de Souza, C. Hotta, J.A. Schlueter, and M. Lang, Nature Mater. {\bf 11}, 755 (2012).
\bibitem{takahashi2006} T. Takahashi, Y. Nogami, and K. Yakushi, J. Phys. Soc. Jpn. {\bf 75}, 051008 (2006).
\bibitem{yakushi2012} K. Yakushi, Crystals {\bf 2}, 1291 (2012).
\bibitem{lunkenheimer2015a} P. Lunkenheimer, B. Hartmann, M. Lang, J. M\"{u}ller, D. Schweitzer, S. Krohns, 
   and A. Loidl, Phys. Rev. B {\bf 91}, 245132 (2015).
\bibitem{sedlmeier2012} K. Sedlmeier, S. Els\"{a}sser, D. Neubauer, R. Beyer, D. Wu, T. Ivek, S. Tomi\'{c}, 
   J.A. Schlueter, and M. Dressel, Phys. Rev. B {\bf 86}, 245103 (2012).
\bibitem{tomic2013} S. Tomi\'{c}, M. Pinteri\'{c}, T. Ivek, K. Sedlmeier, R. Beyer, D. Wu, J.A. Schlueter, 
   D. Schweitzer, and M. Dressel, J. Phys.: Condens. Matter {\bf 25}, 436004 (2013).
\bibitem{lang2014} M. Lang, P. Lunkenheimer, J. M\"{u}ller, A. Loidl, B. Hartmann, N.H. Hoang, E. Gati, 
   H. Schubert, and J.A. Schlueter, IEEE T. Magn. {\bf 50}, 2700107 (2014).
\bibitem{powell2011} B.J. Powell and R.H. McKenzie, Rep. Prog. Phys. {\bf 74}, 056501 (2011).
\bibitem{kino1995a} H. Kino and H. Fukuyama, J. Phys. Soc. Jpn. {\bf 64}, 2726 (1995).
\bibitem{kino1995b} H. Kino and H. Fukuyama, J. Phys. Soc. Jpn. {\bf 64}, 4523 (1995).
\bibitem{kino1996} H. Kino and H. Fukuyama, J. Phys. Soc. Jpn. {\bf 65}, 2158 (1996).
\bibitem{komatsu1996} T. Komatsu, N. Matsukawa, T. Inoue, and G. Saito, J. Phys. Soc. Jpn. {\bf 65}, 1340 
   (1996).
\bibitem{nakamura2009} K. Nakamura, Y. Yoshimoto, T. Kosugi, R. Arita, and M. Imada, J. Phys. Soc. Jpn. 
   {\bf 78}, 083710 (2009).
\bibitem{kandpal2009} H.C. Kandpal, I. Opahle, Y.-Z. Zhang, H.O. Jeschke, and R. Valent\'{i}, 
   Phys. Rev. Lett. {\bf 103}, 067004 (2009).
\bibitem{scriven2012} E.P. Scriven and B.J. Powell Phys. Rev. Lett. {\bf 109}, 097206 (2012).
\bibitem{koretsune2014} T. Koretsune and C. Hotta, Phys. Rev. B {\bf 89}, 045102 (2014).
\bibitem{scriven2009} E. Scriven and B.J. Powell, Phys. Rev. B {\bf 80}, 205107 (2009).
\bibitem{nakamura2012} K. Nakamura, Y. Yoshimoto, and M. Imada, Phys. Rev. {\bf B} 86, 205117 (2012).
\bibitem{shinaoka2012} H. Shinaoka, T. Misawa, K. Nakamura, and M. Imada, J. Phys. Soc. Jpn. {\bf 81}, 
   034701 (2012).
\bibitem{seo2000} H. Seo, J. Phys. Soc. Jpn. {\bf 69}, 805 (2000).
\bibitem{mori2016} T. Mori, Phys. Rev. B {\bf 93}, 245104 (2016).
\bibitem{naka2010} M. Naka and S. Ishihara, J. Phys. Soc. Jpn. {\bf 79}, 063707 (2010).
\bibitem{hotta2010} C. Hotta, Phys. Rev. B {\bf 82}, 241104(R) (2010).
\bibitem{gomi2016} H. Gomi, T.J. Inagaki, and A. Takahashi, Phys. Rev. B {\bf 93}, 035105 (2016).
\bibitem{jawad2010} M. Abdel-Jawad, I. Terasaki, T. Sasaki, N. Yoneyama, N. Kobayashi, Y. Uesu, and 
   C. Hotta, Phys. Rev. B {\bf 82}, 125119 (2010).
\bibitem{gomes2016} N. Gomes, W.W. De Silva, T. Dutta, R.T. Clay, and S. Mazumdar, Phys. Rev. B 
   {\bf 93}, 165110 (2016).
\bibitem{silva2016} W.W. De Silva, N. Gomes, S. Mazumdar, and R.T. Clay, Phys. Rev. B {\bf 93}, 205111 
   (2016).
\bibitem{guterding2016a} D. Guterding, S. Diehl, M. Altmeyer, T. Methfessel, U. Tutsch, H. Schubert, 
   M. Lang, J. M\"uller, M. Huth, H.O. Jeschke, R. Valent\'{i}, M. Jourdan, and H.-J. Elmers, 
   Phys. Rev. Lett. {\bf 116}, 237001 (2016).
\bibitem{guterding2016} D. Guterding, M. Altmeyer, H.O. Jeschke, and R. Valent\'{i}, Phys. Rev. B 
   {\bf 94}, 024515 (2016).
\bibitem{sekine2013} A. Sekine, J. Nasu, and S. Ishihara, Phys. Rev. B {\bf 87}, 085133 (2013).
\bibitem{watanabe2017} H. Watanabe, H. Seo, and S. Yunoki, J. Phys. Soc. Jpn. {\bf 86}, 033703 (2017).
\bibitem{sato2017} N. Sato, T. Watanabe, M. Naka, and S. Ishihara, J. Phys. Soc. Jpn. {\bf 86}, 053701 
   (2017).
\bibitem{okazaki2013} R. Okazaki, Y. Ikemoto, T. Moriwaki, T. Shikama, K. Takahashi, H. Mori, H. Nakaya, 
   T. Sasaki, Y. Yasui, and I. Terasaki, Phys. Rev. Lett. {\bf 111}, 217801 (2013).
\bibitem{kaneko2016b} R. Kaneko, L.F. Tocchio, R. Valent\'{i}, and C. Gros, Phys. Rev. B {\bf 94}, 195111 
   (2016).
\bibitem{private_micha} M. Altmeyer and D. Guterding, private communications.
\bibitem{yokoyama1987} H. Yokoyama  and  H. Shiba.  J. Phys. Soc. Jpn. {\bf 56}, 1490 (1987).
\bibitem{gros1988} C. Gros,  Phys. Rev. B {\bf 38}, 931(R) (1988).
\bibitem{capello2005} M. Capello, F. Becca, M. Fabrizio, S. Sorella, and  E. Tosatti, Phys. Rev. Lett. 
   {\bf 94}, 026406 (2005).
\bibitem{tocchio2016} L.F. Tocchio, F. Arrigoni, S. Sorella, and F. Becca, J. Phys.: Condens. Matter 
   {\bf 28}, 105602 (2016).
\bibitem{tocchio2014} L.F. Tocchio, C. Gros, X.-F. Zhang, and S. Eggert, Phys. Rev. Lett. {\bf 113}, 
   246405 (2014).
\bibitem{kaneko2016a} R. Kaneko, L.F. Tocchio, R. Valent\'{i}, F. Becca, and C. Gros, Phys. Rev. B 
   {\bf 93}, 125127 (2016).
\bibitem{feynman1954} R.P. Feynman, Phys. Rev. {\bf 94}, 262 (1954).
\bibitem{tocchio2011} L.F. Tocchio, F. Becca, and C. Gros, Phys. Rev. B {\bf 83}, 195138 (2011).
\bibitem{hotta2006} C. Hotta and N. Furukawa, Phys. Rev. B {\bf 74}, 193107 (2006).
\bibitem{fevrier2015} C. F\'{e}vrier, S. Fratini, and A. Ralko, Phys. Rev. B {\bf 91}, 245111 (2015).
\bibitem{watanabe2003} M. Watanabe, Y. Noda, Y. Nogami, and H. Mori, Synth. Met. {\bf 135}, 665 (2003).
\bibitem{kakiuchi2007} T. Kakiuchi, Y. Wakabayashi, H. Sawa, T. Takahashi, and T. Nakamura, J. Phys. Soc. Jpn. 
   {\bf 76}, 113702 (2007).
\end{thebibliography}
\end{document}